\documentclass{ifacconf}

\usepackage{graphicx}      
\usepackage{natbib}        

\usepackage{tikz}
\usetikzlibrary{positioning, tikzmark, calc, arrows.meta}
\usepackage{pgfplots}
\pgfplotsset{compat=1.17}
\usepackage{graphicx}
\usepackage{lipsum}

\usepackage{mathtools}

\usepackage{amsmath}
\usepackage{amsfonts}
\usepackage{amssymb}
\usepackage{mathrsfs}
\usepackage{xspace}
\usepackage{xparse}

\usepackage{fancyhdr}

\begin{document}
\begin{frontmatter}

\title{This work has been submitted to IFAC for possible publication}

\title{Characterizing Within-Driver Variability in Driving Dynamics During Obstacle Avoidance Maneuvers\thanksref{footnoteinfo}}

\thanks[footnoteinfo]{This material is based upon work supported by the National Science Foundation, under Grant Numbers CNS-1836900 and CNS-1836952. Any opinions, findings, and conclusions or recommendations expressed in this material are those of the authors and do not necessarily reflect the views of the National Science Foundation. 
}

\author[First]{Kendric R. Ortiz} 
\author[First]{Adam J. Thorpe} 
\author[Second]{AnaMaria Perez}
\author[Third]{Maya Luster}
\author[Third]{Brandon J. Pitts}
\author[First]{Meeko Oishi}

\address[First]{University of New Mexico, Elec. and Comp. Eng., Albuquerque, NM 87131, USA (e-mail: kendric, ajthor,  oishi@unm.edu)}

\address[Second]{Harvard University, Mathematics, Cambridge, MA 02138, USA (e-mail: anamariaperez@college.harvard.edu )}
  
\address[Third]{Purdue University, Industrial Eng., West Lafayette, IN 47907, USA (e-mail: mluster, bjpitts@purdue.edu)}

\begin{abstract}                
Variability in human response creates non-trivial challenges for modeling and control of human-automation systems.  
As autonomy becomes pervasive,
methods that can accommodate human variability will become paramount, to ensure efficiency, safety, and high levels of performance.  We propose an easily computable modeling framework which takes advantage of a metric to assess variability in individual human response in a dynamic task that subjects repeat over several trials.  Our approach is based in a transformation of observed trajectories to a reproducing kernel Hilbert space, which captures variability in human response as a distribution embedded within the Hilbert space.  We evaluate the similarity across responses via the maximum mean discrepancy, which measures the distance between distributions within the Hilbert space.  We apply this metric to a difficult driving task designed to elucidate differences across subjects.  We conducted a pilot study with 6 subjects in an advanced driving simulator, in which subjects were tasked with
collision avoidance of an obstacle in the middle of the road, around a blind corner, in a nighttime scenario, while steering only with the non-dominant hand.  
\end{abstract}


\end{frontmatter}

\section{Introduction}
Designing autonomy to be human-centric 
is paramount for effective and safe human-automation interaction.  Indeed, autonomy that is not responsive to the human in the loop can create conditions ripe for ``misuse, disuse, [and] abuse'' of autonomy \citep{Parasuraman97}.  Customization to individual subjects is one way to make autonomy human-centric \citep{Tomizuka2020RAL,Thropp2018THMS}, and has the potential to 
significantly impact performance and reliability \citep{Abbink2016TH,Srinivasa2012ROMAN} in a manner that can improve human-automation interaction in a wide range of applications.  

However, there is a dearth of methods and tools that can effectively characterize heterogeneity across subjects, as well as variability of a given subject over time.  
Historically, models that capture sufficient detail to describe an individual subject have been computationally intractable \citep{ACT-R}, or involve non-reproducible modeling processes.  
Tools that can capture relevant metrics and can be tailored to an individual, in a repeatable manner, are foundational to the customization and personalization of autonomy for cyber-physical systems (CPS).
{\em In this paper, we propose the use of data-driven techniques to assess variability in individual driver response, based on a metric that captures differences between distributions within a reproducing kernel Hilbert space.}  We focus on a driving scenario with naturalistic decision making that confounds traditional control theoretic modeling \citep{wickens2021engineering,weir1970dynamics,hess}.  


In driving assistance systems, customization to driver preferences, characteristics, and behaviors has been shown to be important for system effectiveness \citep{Chen2020TIV, Hasenjager2017}.  In collision avoidance systems, researchers have investigated data-driven approaches to characterize preferred driver risk \citep{Muehlfeld2013, Lefevre2015}, and to trigger alerting systems \citep{Wang2016, Inata2008} that engage assistive control.  Customization impacts controller design in a variety of contexts, and has been investigated in several applications, including fault tolerance \citep{Chen2020TIV}, multi-objective car-following \citep{MPC2020TVT}, and lane-keeping \citep{Rath2019IET, Noto2011}, in which the resulting trajectories must be deemed acceptable to an individual driver.  Other work has focused on reducing potential conflict between the driver and the autonomous system in a shared control framework \citep{Abbink2014SMC} and via potential fields \citep{Noto2011}. 

\begin{figure}
    \centering
    \label{fig:photo}
    \begin{tikzpicture}
    \node (I) at (0, 0) {\includegraphics[width = 0.9\columnwidth]{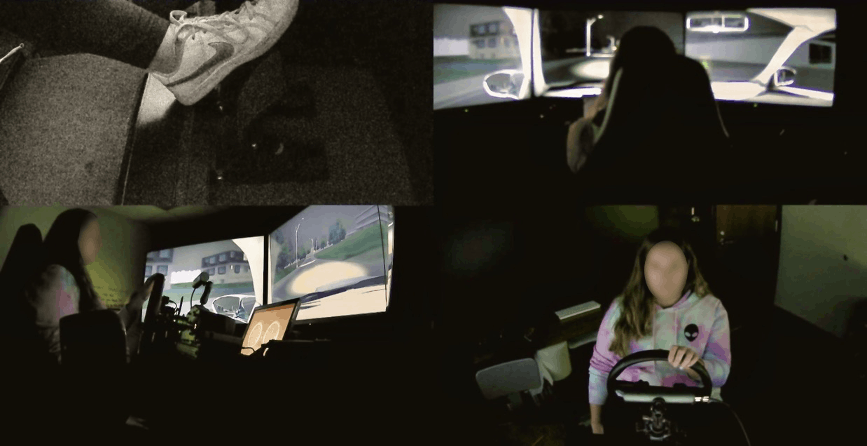}};
    \fill[red] (2.13,-0.45) circle (5pt);
    \end{tikzpicture}
    \caption{Participants engaged in a collision avoidance scenario in an advanced driving simulator, steering one-handed, with their non-dominant hand, at night, around an obstacle obscured by a curve in the road.}
\end{figure}

We focus on characterization of within-driver variability in a challenging simulated driving scenario (Figure \ref{fig:photo}), that is, changes in a single driver's performance over repeated trials, in a scenario carefully constructed to test the subject's ability to adapt to an unfamiliar experience \citep{Luster2021HFES}.  
This pilot experiment took place in an advanced driving simulator, with a reasonable degree of realism.  Gathered data included not only simulated vehicle trajectories, but also eye tracking data.  

Our approach to characterize variability is based in a non-parametric machine learning technique, kernel distribution embedding, which exploits properties of reproducing kernel Hilbert spaces.  This method enables a data-driven analysis of observed trajectories in a manner that minimizes bias due to tuning, a lack of modeling assumptions, and other heuristics.  We project entire trajectories into the Hilbert space, then use the maximum mean discrepancy (MMD) to characterize the distance between the distributions of trajectories in the RKHS. 
{\em Our main contribution is a methodological assessment of variability in driving response, via kernel distribution embeddings, and a characterization of its relationship to eye tracking data.}




\section{Experimental Setup}
\label{section: experimental setup}

\subsection{Driving scenario}
\label{section: experiment setup}

\begin{figure}
    \centering
    \includegraphics[height=1.5in]{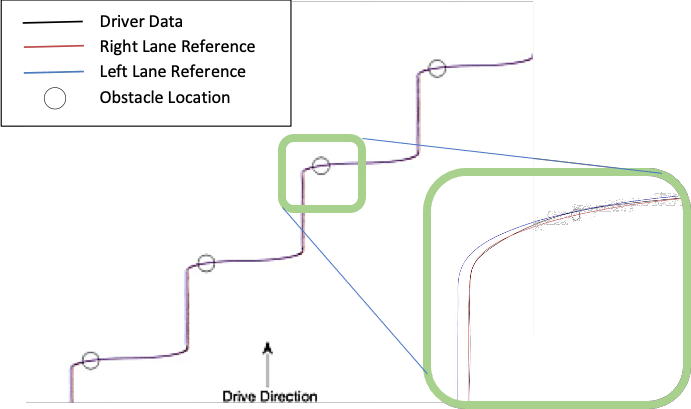}
    \caption{ Experimental driving simulator set up, where the obstacle used for avoidance is an old car tire located in the center of the lane at each curve in the road.}
    \label{fig: experiment trial}
\end{figure}

A novel driving scenario was constructed as part of a pilot study at Purdue University, approved under IRB No. 1905022220.  
The driving scenario consisted of an obstacle avoidance task, in which the participant was to drive in a nighttime setting on a two-lane rural-type highway (Figure \ref{fig:photo}), and avoid any roadway obstacles that might be present.  The driving environment consisted of four S-curve shaped segments, and four straightaways that preceded each S-curve segment (Figure \ref{fig: experiment trial}).  An obstacle (an old vehicle tire) was placed at the beginning of each S-curve, 
such that it was not visible to the participant until they rounded the curve.  Streetlights were present along both sides of the road for the duration of the drive. 
There were no other vehicles in the scenario, i.e., no oncoming traffic, and no leading or trailing vehicles. 

\begin{figure}
    \centering
    \includegraphics[width=3.0in]{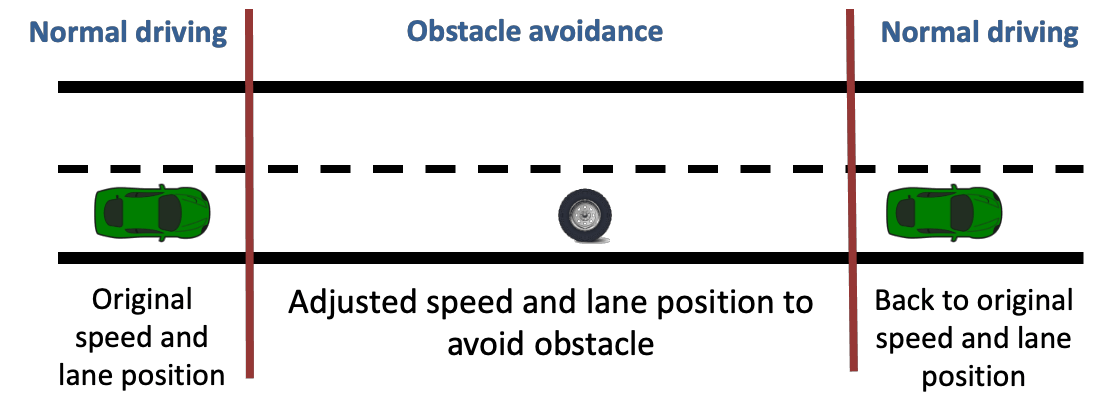}
    \caption{Participants were tasked with avoiding an old tire, which required deviating out of their lane, then returning to their lane as quickly as possible.}
    \label{fig:driving_cartoon}
\end{figure}

Participants were asked to drive one-handed, with their non-dominant hand, throughout the entire experiment, to increase the level of difficulty in an otherwise familiar task.  
Each participant was given the same set of instructions to complete the driving task (Figure \ref{fig:driving_cartoon}): (1) practice safe driving (i.e., stay within the lane boundaries), (2) drive at a constant speed of 45 mph throughout the session, and (3) if an obstacle was encountered, avoid the obstacle by moving into the opposite lane as quickly as possible; once cleared, move back into the original lane and resume a constant speed of 45 mph. 
Participants were first asked to complete a practice drive during daytime on an open highway to familiarize themselves with the driving simulator and task, before engaging in the driving task.

\subsection{Collected data and participant demographics}

\begin{figure}
    \centering
    \includegraphics[width=\columnwidth]{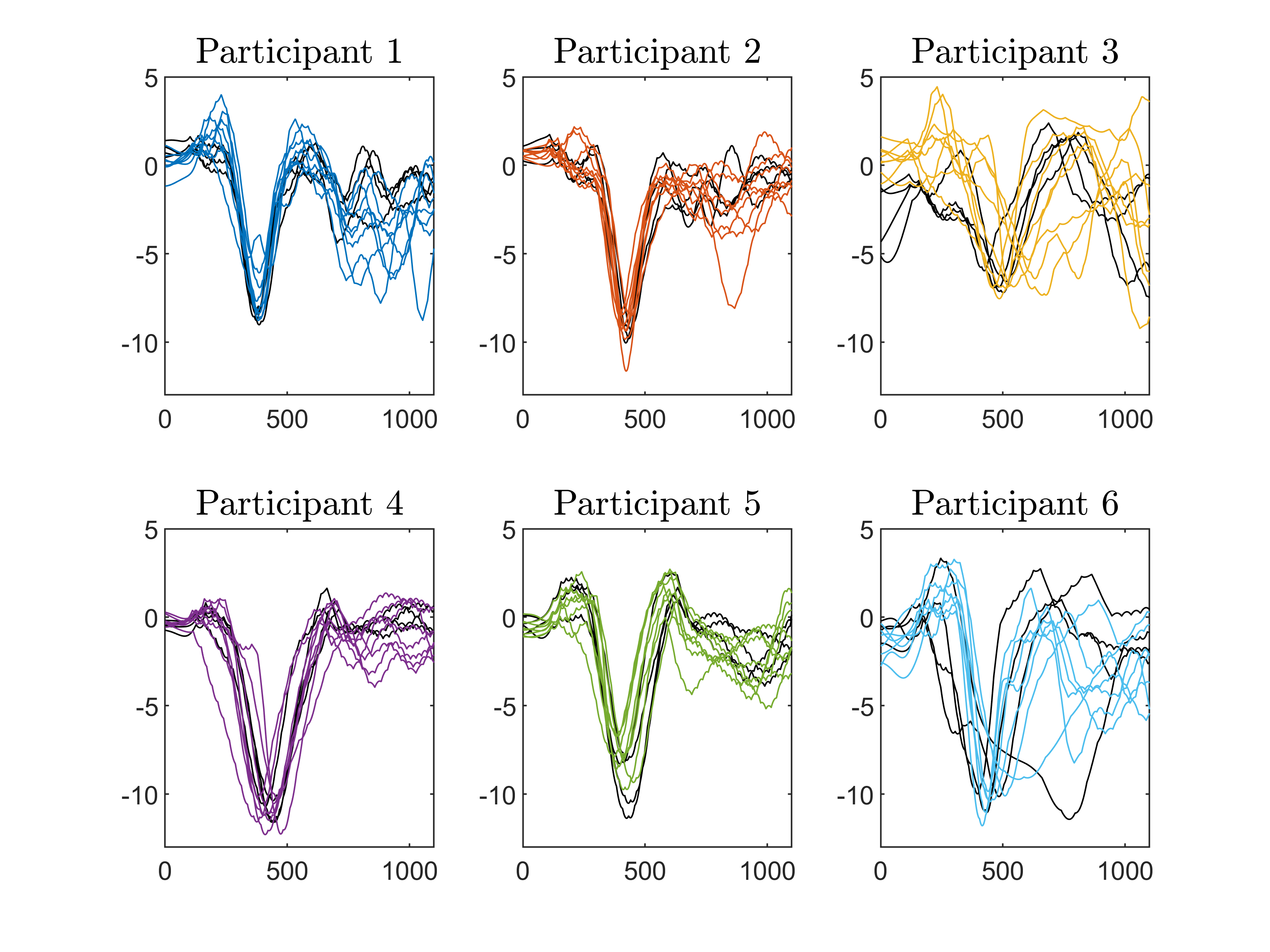}
    \caption{Each participant generated 12 trajectories.  Vehicle position with respect to the centerline is separated for the first 8 trajectories (color) and the last 4 (black).}
    \label{fig: isolated trajectories}
\end{figure}

A total of 6 participants completed this pilot experiment.  
The experiment consisted of three trials, each of which contained four consecutive S-curves; hence, we obtain a total of 12 trajectories for each participant.  
Trajectories describe position with respect to the centerline of the road (Figure \ref{fig: isolated trajectories}), velocity, and heading of the vehicle.  
Each trial lasted approximately four minutes.

Eye-tracking data was collected as described in \cite{Luster2021HFES} for a 7 second duration before each obstacle.  An area of interest demarcated an area within the vehicle's lane prior to each obstacle, and average fixation duration was calculated as the average time that the participant was fixated on the area of interest.

Three of the 6 participants were male, and 3 were female; all were senior students in engineering. Average participant age was 21.33 years, with a standard deviation of 0.82 years.  Driving experience for each of the participants is summarized in Table \ref{tab: Driving_experience}; the average years of driving experience was 4.17 years.  

\begin{table}
    \centering
    \begin{tabular}{|c|c|}
         \hline  Participant number & Driving experience  \\ 
         \hline 
         1 &  6 years\\
         2 & 3 years \\
         3 & 1 year \\
         4 & 6 years \\
         5 & 2 years \\
         6 & 7 years  \\
         \hline
    \end{tabular}
    \caption{Participants' driving experience}
    \label{tab: Driving_experience}
\end{table}

\section{Data Driven Models of Driver-In-The-Loop CPS}
\label{sec: System model and MMD}

\subsection{Challenges in modeling human-in-the-loop CPS}

Human-in-the-loop systems pose a significant challenge for modeling and control, particularly in configurations in which human response is hard to predict, such as learning scenarios, or other circumstances which invoke naturalistic decision making \citep{wickens2021engineering}.
While a variety of models, of varying degrees of complexity and veracity, have been explored in driving tasks, most are effective under highly constrained scenarios, for which human operators have extensively trained and practiced \citep{weir1970dynamics, sheridan1992telerobotics}. For the learning scenario we consider, an intentional lack of training and practice confounds these approaches. Hence, one significant challenge is the lack of known mathematical models of the human’s inputs to the system.

Presumptions of accurate prior knowledge of the dynamics, as well as of the uncertainty due to the human input, are unrealistic, and could lead to inaccurate assessments, as well as an inability to distinguish features important for within-driver variability. Further, assumptions regarding the stochastic properties of the system (such as Gaussianity or the Markov property) may simply be inaccurate. Lastly, heterogeneity of human driving behavior means there is likely no single characterization or model which is applicable to all subjects, particularly during the learning phase, as drivers gain experience with the driving task.

To sidestep these difficult and potentially unsolvable modeling challenges, we take an empirical, data-driven approach.  We model human driver behavior using a statistical learning technique, kernel embeddings of distributions.
Although this approach neglects knowledge of the vehicle dynamics, it captures statistical information about the complex interconnections between human inputs and the underlying dynamical system.
This approach constructs an implicit representation of the probability distribution of an (unknown) stochastic process in a reproducing kernel Hilbert space (RKHS), which has two key advantages: 1) it is nonparametric, meaning it does not impose prior assumptions on the structure of the uncertainty; 2) it admits a distance metric, the maximum mean discrepancy (MMD), which characterizes the similarity between two distributions.


\subsection{Modeling via stochastic processes}

We consider the driver-in-the-loop system to be a general stochastic process \citep{Cinlar2011}, 
\begin{equation}
    \label{eqn: stochastic process}
    X = \lbrace X_{t} \mid t \in \mathcal{T} \rbrace,
\end{equation}
such that for each $t \in \mathcal{T}$, an index set representing time, $X_{t}$ is a random variable that takes values in the measurable space $(\mathcal{X}, \mathscr{B}(\mathcal{X}))$, where $\mathscr{B}(\mathcal{X})$ is the Borel $\sigma$-algebra of $\mathcal{X}$ generated by the set of all open subsets of $\mathcal{X}$.
The stochastic process $X$ is a random variable over the space $\Xi = \mathcal{X}^{\mathcal{T}}$, which is the product space of $\mathcal{X}$ over $\mathcal{T}$. We denote the probability distribution of $X$ as $\mathbb{P}$, and realizations of $X$ are called trajectories, denoted $\xi \in \Xi$.

Using data collected from the driving experiment, we seek to compute an empirical representation of the distribution of $X$ \eqref{eqn: stochastic process} as an element in an RKHS. 

\begin{defn}[RKHS, \citealp{Aronszajn1950}]
    A Hilbert space of functions from $\Xi$ to $\mathbb{R}$ is a reproducing kernel Hilbert space (RKHS) if there exists a positive definite function $k : \Xi \times \Xi \to \mathbb{R}$, called the kernel, that satisfies the following properties:
    \begin{enumerate}
        \item 
        For every $\xi \in \Xi$, $k(\xi, \cdot) \in \mathscr{H}$, and 
        \item 
        (Reproducing property): For every $\xi \in \Xi$ and $f \in \mathscr{H}$, $f(\xi) = \langle f, k(\xi, \cdot) \rangle_{\mathscr{H}}$.
    \end{enumerate}
\end{defn}

A common choice of kernel function is the Gaussian kernel, $k(\xi, \xi') = \exp(- \lVert \xi - \xi' \rVert^{2}/2\sigma^{2})$.
Furthermore, the RKHS is equipped with the inner product $\langle \cdot, \cdot \rangle_{\mathscr{H}}$ and the implicitly defined norm $\lVert \cdot \rVert_{\mathscr{H}}$.

At the heart of our approach is an element $m_{\mathbb{P}} \in \mathscr{H}$, the kernel distribution embedding of $\mathbb{P}$ \citep{smola2007hilbert}, defined as
\begin{equation}
    \label{eqn: kernel mean embedding}
    m_{\mathbb{P}} = \int_{\Xi} k(y, \cdot) \mathbb{P}(\mathrm{d} y),
\end{equation}
where $\mathbb{E}[f(X)] = \langle f, m_{\mathbb{P}} \rangle_{\mathscr{H}}$ by the reproducing property.
This representation allows us to capture the probability distribution of $X$ as an element of the RKHS. 

However, in practice, we cannot compute
\eqref{eqn: kernel mean embedding} directly, since the distribution $\mathbb{P}$ of $X$ is unknown.
Hence, following \cite{smola2007hilbert}, we compute an empirical estimate $\widehat{m}_{\mathbb{P}}$ 
from data $\mathcal{D} = \lbrace \xi^{i} \rbrace_{i=1}^{M}$, collected i.i.d. from $\mathbb{P}$, such that
\begin{equation}
    \widehat{m}_{\mathbb{P}} = \frac{1}{M} \sum_{i=1}^{M} k(\xi^{i}, \cdot).
    \label{eqn: empirical mean embedding}
\end{equation}
{In effect, this enables empirical estimation of the probability distribution of the driver-in-the-loop system as an element of the RKHS.}

\subsection{Maximum mean discrepancy}

One particularly useful property of the RKHS facilitates analysis of the similarity of functions and distributions. The implicit distance metric in the RKHS $\mathscr{H}$, $\lVert f - g \rVert_{\mathscr{H}}$, where $\lVert f - g \rVert_{\mathscr{H}} = 0$ implies $f = g$, can be used to 
measure the similarity of two distributions $\mathbb{P}, \mathbb{Q}$ on $\Xi$ via their embeddings, $m_{\mathbb{P}}$ and $m_{\mathbb{Q}}$, 
\begin{equation}
    \label{eqn: MMD}
    \mathrm{MMD}(\mathbb{P}, \mathbb{Q}) = \lVert m_{\mathbb{P}} - m_{\mathbb{Q}} \rVert_{\mathscr{H}}.
\end{equation}
This is the {\em maximum mean discrepancy} (MMD) \citep{Gretton2012}. 

Importantly, if the kernel is \emph{characteristic} \citep{sriperumbudur2010}, then the embedding is unique, meaning it captures all statistical features of the underlying distribution and is injective \citep[Theorem~1]{smola2007hilbert}. 
In other words, if $k$ is characteristic,
then for any two distributions $\mathbb{P}$ and $ \mathbb{Q}$,  $\lVert m_{\mathbb{P}} - m_{\mathbb{Q}} \rVert_{\mathscr{H}} = 0$ if and only if $\mathbb{P} = \mathbb{Q}$.
Many common choices of kernels are characteristic, including the Gaussian kernel that we employ.

Using the empirical estimate \eqref{eqn: empirical mean embedding}, we can compute an estimate of \eqref{eqn: MMD} as
\begin{equation}
    \widehat{\text{MMD}}(\mathbb{P},\mathbb{Q}) 
    =\left\|\frac{1}{M}\sum^M_{i=1}k(\xi^{i}, \cdot)-\frac{1}{N}\sum^N_{i=1}k(z^{i}, \cdot)\right\|_{\mathscr{H}}.
    \label{eqn: empirical MMD}
\end{equation}
Thus, we can readily employ the MMD to assess the similarity between empirical estimates of distributions.

We note that although the maximum mean discrepancy is a type of integral probability metric \citep{Sriperumduhur2012}
(such as the KL divergence, or total variation divergence), it does not require density estimation, which can be computationally expensive \citep{sriperumbudur2010, Gretton2012}. This computational ease makes the MMD a good candidate for near-run time applications for the purpose of control. 
Additionally, alternative approaches, such as those based in moment matching, may not only be difficult to employ with limited data, but also computationally expensive, as a high number of moments may be required to ensure sufficient accuracy.

\section{Analysis and Results}
\label{section: Analysis and Results}

\subsection{Computing empirical estimates of embeddings}
\label{subsec: MMD Analysis}

We presume the observed data is of the form
\begin{equation}
    \mathcal{D} = \lbrace \xi^{i} \rbrace_{i=1}^{M}
\end{equation}
where $M \in \mathbb{N}$ denotes the total number of trajectories in the data set $\mathcal{D}$,
and each trajectory $\xi^{i} \in \mathcal{D}$ is a sequence of state vectors indexed by time,
\begin{equation}
    \xi^{i} = \lbrace x_{0}^{i}, x_{1}^{i}, \ldots, x_{N}^{i} \rbrace
\end{equation}
where $x_{t}^{i} \in \mathcal{X}$ for all $i = 1, \ldots, M$ and $t = 0, 1, \ldots, N$.
We denote $\mathcal{D}_{j}$ to be the set of all trajectories in $\mathcal{D}$ corresponding to the $j^{\rm th}$ subject. 
We assume that the trajectories of each participant $j$ are drawn i.i.d. from stochastic process $X^{j}$ with unknown probability distribution $\mathbb{P}_{j}$, where the stochasticity is due to the human-in-the-loop operator.

Hence for each subject $j$, with $M = 12$, we have data $\mathcal{D}_{j} = \lbrace \xi_{j}^{1}, \xi_{j}^{2}, \ldots, \xi_{j}^{12} \rbrace$, with trajectories $\xi_j^i \in \mathbb R^{3N}$ over $N = 1100$ time steps, at a sampling rate of 60 Hz.  



We first separate the first eight trajectories ($\xi_j^1, \cdots, \xi_j^{8}$) from the last four trajectories ($\xi_j^9, \cdots, \xi_j^{12})$, which describe the {\em final distribution}.  
We seek to compare each of the first eight trajectories with the final distribution to assess consistency of driving performance over time.  


We first compute the empirical kernel distribution embedding of the final distribution from (\ref{eqn: empirical mean embedding}), as
\begin{equation}
    \widehat{m}_{\mathbb{P}_{j}} = \frac{1}{4} \sum_{i=9}^{12} k(\xi_{j}^{i}, \cdot)
\label{eq:final}
\end{equation}
for $\xi_{j}^{i} \in \mathcal{D}_{j}$. 
We used a Gaussian kernel with bandwidth parameter $\sigma = 32$, chosen via cross validation to ensure that the MMD can distinguish between distributions.
%

We then compute the maximum mean discrepancy for each of the first 8 trajectories with respect to (\ref{eq:final}), by evaluating (\ref{eqn: empirical MMD}) for each, as
\begin{equation}
    \widehat{\textnormal{MMD}}({\xi_{j}^{i}}, \mathbb{P}_{j}) = \lVert k(\xi_{j}^{i}, \cdot) - \widehat{m}_{\mathbb{P}_{j}} \rVert_{\mathscr{H}}.
\end{equation}

Because the Gaussian kernel is characteristic, 
the MMD will decrease as trajectories become more consistent with the final distribution, and the MMD will increase when trajectories became more dissimilar.  

\begin{figure}[t]
    \centering
    \includegraphics[width = .75\linewidth]{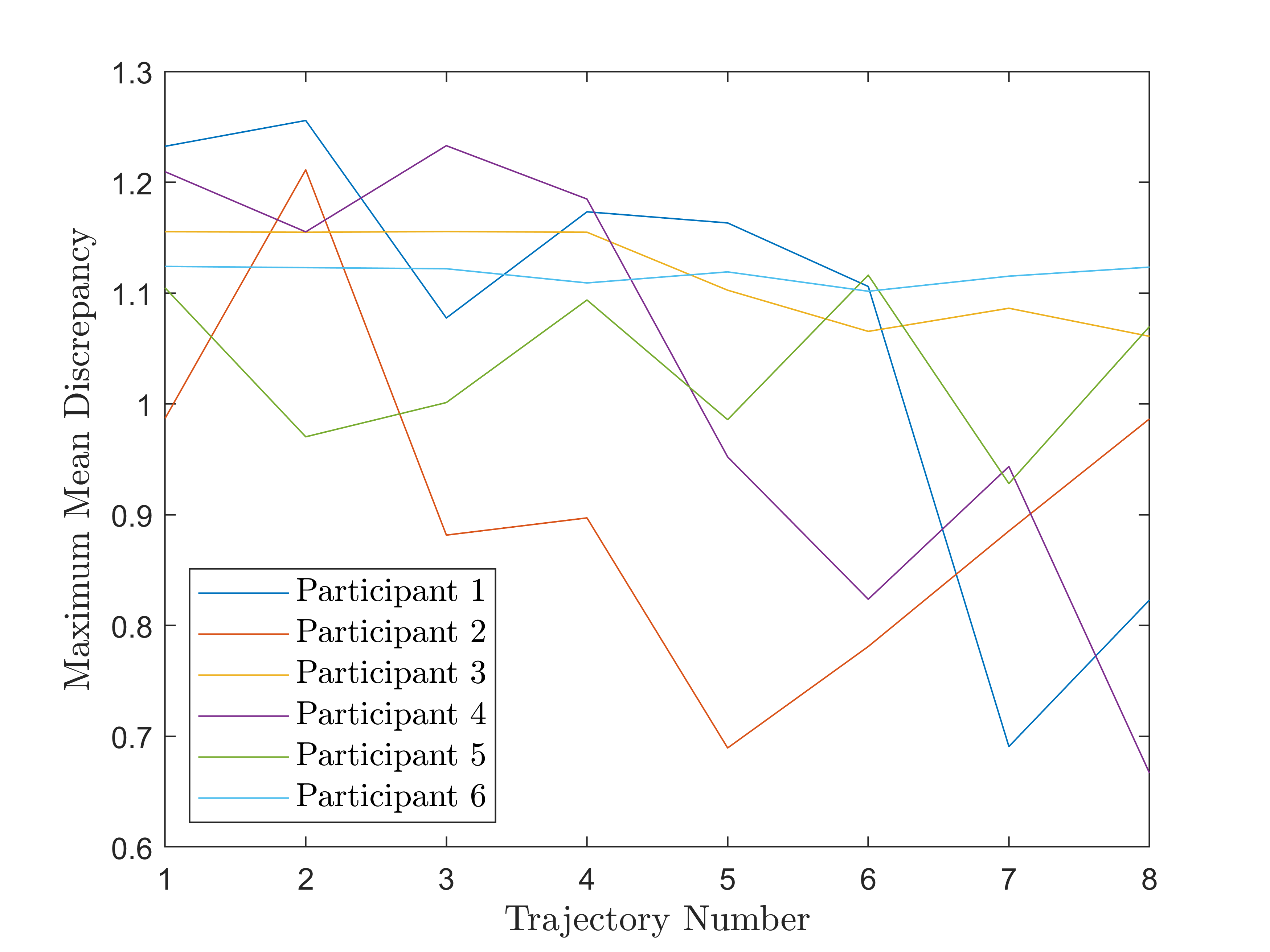}
    \caption{Maximum mean discrepancy for all 6 participants in as a function of time. Participants 1, 3, and 4 show decreasing trends; participants 5 and 6 show variability but no overall change; participant 2 shows an initial decrease over the first four trajectories, followed by an increase over the next four trajectories.}
    \label{fig: all_subjs}
\end{figure}

\subsection{Trajectory analysis via maximum mean discrepancy}

As shown in Figure \ref{fig: all_subjs}, there is considerable heterogeneity in trends across subjects. 
Participants 1, 3, and 4 show a decreasing trend in MMD as trajectories progress, indicating that their performance became more consistent with the final distribution as the participants gained familiarity with the obstacle avoidance task.  We interpret this trend as evidence of learning, because of the improvements in performance over time.  

Participants 5 and 6 show no major increases or decreases in MMD over the duration of the entire experiment.  Participant 5 has a fairly consistent MMD, whereas Participant 6 shows considerable variability, with MMD values that range between 0.9 and 1.1.  In both cases, we infer that driving performance was essentially unchanged in its variability over the duration of the first two trials (i.e., first 8 trajectories) of the experiment.  There are several potential explanations for this finding, including a ceiling effect (i.e., maximum performance has been reached with little to no room for improvement), a lack of environmental stimuli (e.g., no on-coming, opposing traffic) to encourage improvement, no motivation for enhancing performance given the absence of experiment incentives, and/or behavior being driven by one’s own expectation and interpretation of the goal(s) of the study. 

Participant 2 shows a completely different trend: this participant has a decrease in MMD over the first four trajectories, but an increase in MMD over the next four trajectories.  We believe that this is due to two primary factors: 1) a loss of attention that was informally observed during the experiment, and 2) a poor understanding of the instructions, evident, in part, by the participant's lack of focus as the experiment progressed, as well as the fact that the participant employed two hands (contrary to instructions) for the first four trajectories. 

\subsection{Correlation with eye-tracking analysis}
\begin{figure}
    \centering
    \includegraphics[width = .75\columnwidth, keepaspectratio = true]{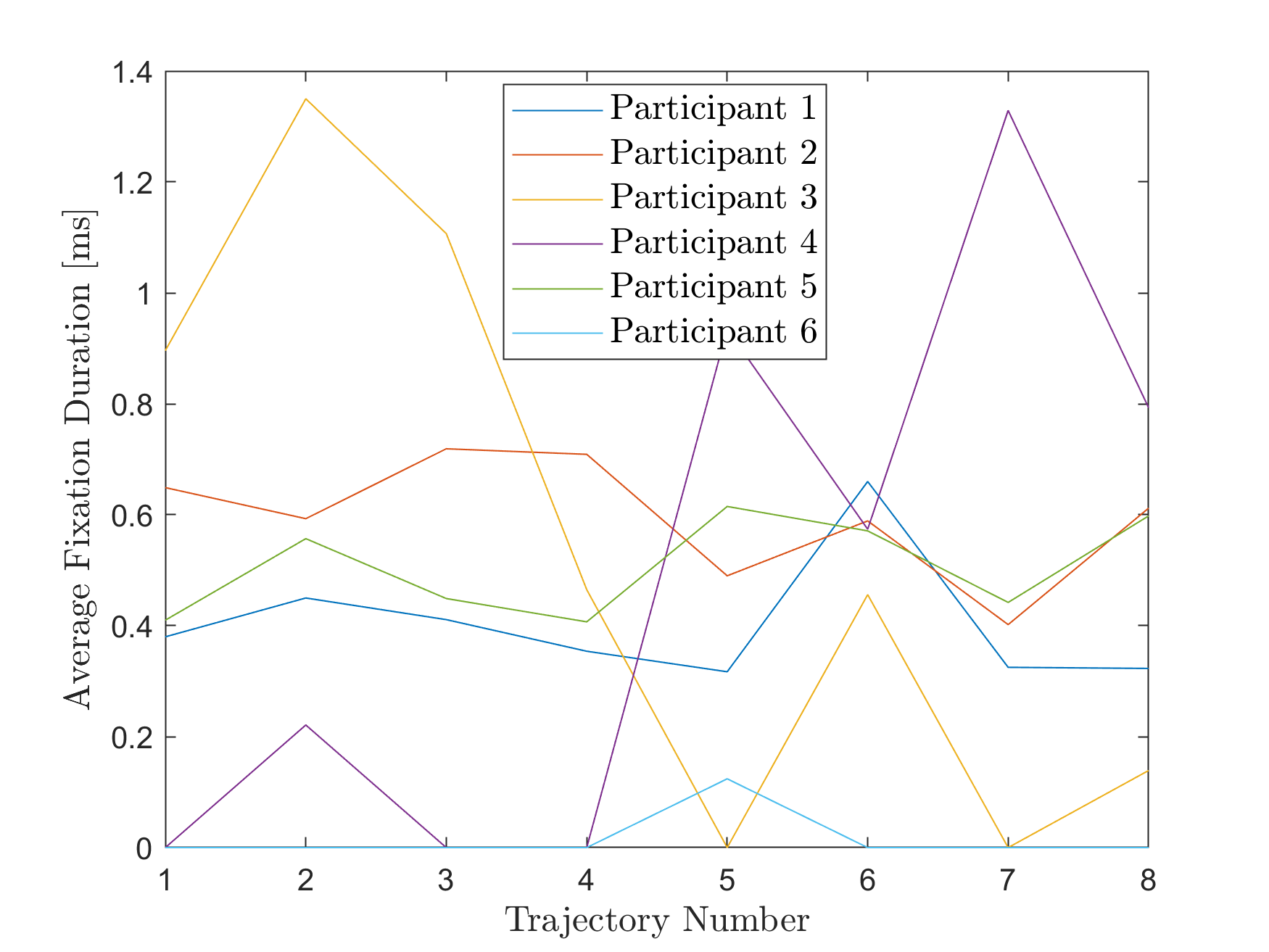}
    \caption{Average fixation duration in the 7 seconds before the obstacle, for the first 8 trajectories for all 6 participants.}
    \label{fig: eye-tracking}
\end{figure}

\begin{figure}
    \centering
    \includegraphics[width = .75\columnwidth, keepaspectratio = true]{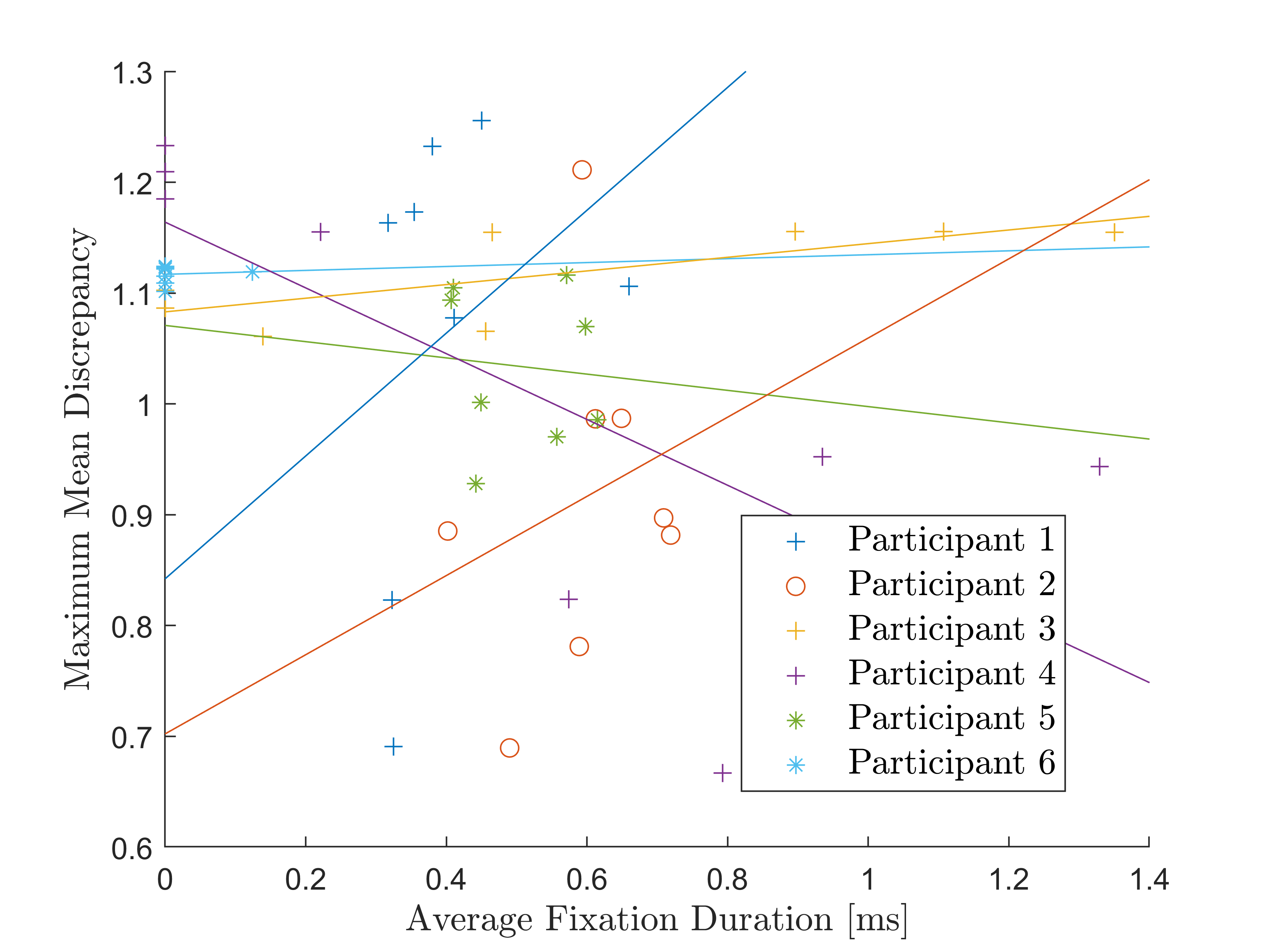}
    \caption{The relationship between maximum mean discrepancy and average fixation duration shows different trends for participants who exhibited learning (`+') and those who did not (`*').  Linear regression indicates small absolute slopes for those participants who did not show improvements in MMD, and larger absolute values of slopes for those who did.}
    \label{fig: Avgerage fixation vs MMD}
\end{figure}

Since eye tracking analysis can provide insights into attention allocation, we also consider the relationship between driving performance (indicated by MMD) and eye behavior, i.e., average fixation duration, a measure of the amount of time a person visually engages with particular area of interest. 
We consider the average fixation duration in the 7 seconds before each obstacle for all 6 participants (Figure \ref{fig: eye-tracking}), and construct a linear regression with respect to MMD (Figure \ref{fig: Avgerage fixation vs MMD}).

Participants that exhibit learning (i.e., those who have improvements in driving performance as the experiment progresses) have divergent trends (Figure \ref{fig: Avgerage fixation vs MMD}).  This possibly indicates different strategies employed by participants.  
For example, Participant 1 showed an decrease in attention over time, which is consistent with the hypothesis in \citep{Luster2021HFES}, that as learning occurs, a shorter fixation duration is needed to complete an identical task, given that familiarity is built based on previous trials.  In contrast, Participant 4 shows increased attention over time, possibly due to their belief that more attention was required to achieve improved performance.
We believe that the difference in trends between Participant 3 and Participant 4 may be evidence of different learning strategies.  The former may show less effort required to achieve improved performance, whereas the latter may be evidence of more effort at participants' discretion.  

Participants 5 and 6, who do not exhibit clear trends in learning, show relatively static linear fit trends in eye-tracking, as well.  In essence, both driving performance and attention appear to be unchanged for the duration of the experiment.
We note that while Participant 3 also has a relatively constant eye-tracking linear fit (see Figure \ref{fig: Avgerage fixation vs MMD}), their performance is more consistent with Participants 1 and 4 than it is with Participants 5 and 6 (as shown in Figure \ref{fig: all_subjs}).  
This suggests that Participant 3 might have not adopted any particular attentional strategies in successive trials, which by chance, resulted in marginally poorer driving performance over time. However, confirmation of this hypothesis would require additional participants beyond this pilot study, and partitioning them into groups based on distinctive similarities.

Lastly, Participant 2 shows a trend in average fixation duration that is similar to that of Participants 5 and 6, and is consistent with the lack of attention informally observed for this participant.  We note that the strong positive slope in Figure \ref{fig: Avgerage fixation vs MMD} may be difficult to interpret because it involves data from essentially two different stochastic processes (two-handed driving in the first four trajectories, and one-handed driving in the next four trajectories).  Separating the data results in a slope of -2.51 for the first four trajectories (comparable to that of Participant 4), and a slope of +0.31 for the next four trajectories (comparable to that of Participants 5 and 6.  We believe this shows evidence of learning in the first four trials, and disengagement in the second four trials. 

\section{Implications for Control}
\label{section: control implications}

One of the major contributions of this work is the quantitative assessment of 1) within-subject variability in a difficult driving task, as well as 2) heterogeneity across subjects in those trends.  The variability and heterogeneity demonstrated here has potential for significant impact in control and in autonomous systems.  Specifically, a one-size-fits all control could have a detrimental impact on some participants.  Consider a controller designed to help improve performance of novices, by helping them learn the desired task more quickly than they might learn it on their own.  For example, a controller designed to intervene in response to increasing fixation durations could actually impair the learning process for participants whose learning strategies involve increased attention at their performance improves.  Similarly problematic phenomena could arise in driver assistance systems designed to detect exceptions to anticipated phenomena, for example.  

We maintain that development of future autonomous systems will require accurate assessments of individual performance, customized not only to the individual, but also responsive to the operational context of human CPS. 




\section{Conclusions} \label{section: conculsion}

We provide a data-driven approach to assess variability in a difficult driving scenario.  Because mathematical models are scarce for naturalistic scenarios such as those that involve learning, our approach exploits kernel distribution embeddings, which limit potential inaccuracies posed by incorrect model assumptions.  We characterize variability in the driving task, and show potentially different learning strategies for those participants that show evidence of learning through driving performance.  Based on our preliminary work here, we propose that that characterization of heterogeneity is indeed important for design of autonomy that is customizable and truly human-centric, and that accounting for individual variability is important for future work in customizable control of human CPS.  







\bibliography{IEEEabrv, shortIEEE, bibliography}

\end{document}